# A platform for time-resolved scanning Kerr microscopy in the near-field

Paul S. Keatley, Thomas H. J. Loughran, Euan Hendry, William L. Barnes, and Robert J. Hicken

*Department of Physics and Astronomy, University of Exeter, EXETER, EX4 4QL, UK*

Jeffrey R. Childress and Jordan A. Katine

*San Jose Research Center, HGST, a Western Digital Company, San Jose, CA 95135, USA*

Time-resolved scanning Kerr microscopy (TRSKM) is a powerful technique for the investigation of picosecond magnetization dynamics at sub-micron length scales by means of the magneto-optical Kerr effect (MOKE). The spatial resolution of conventional (focused) Kerr microscopy using a microscope objective lens is determined by the optical diffraction limit so that the nanoscale character of the magnetization dynamics is lost. Here we present a platform to overcome this limitation by means of a near-field TRSKM that incorporates an atomic force microscope (AFM) with optical access to a metallic AFM probe with a nanoscale aperture at its tip. We demonstrate the near-field capability of the instrument through the comparison of time-resolved polar Kerr images of magnetization dynamics within a microscale NiFe rectangle acquired using both near-field and focused TRSKM techniques at a wavelength of 800 nm. The flux-closure domain state of the in-plane equilibrium magnetization provided the maximum possible dynamic polar Kerr contrast across the central domain wall, and enabled an assessment of the magneto-optical spatial resolution of each technique. Line profiles extracted from the Kerr images demonstrate that the near-field spatial resolution was enhanced with respect to that of the focused Kerr images. Furthermore, the near-field polar Kerr signal (~1 mdeg) was more than half that of the focused Kerr signal, despite the potential loss of probe light due to internal reflections within the AFM tip. We have confirmed the near-field operation by exploring the influence of the tip-sample separation, and have determined the spatial resolution to be ~550 nm for an aperture with a sub-wavelength diameter of 400 nm. The spatial resolution of the near-field TRSKM was in good agreement with finite element modelling of the aperture. Large amplitude electric field along regions of the modelled aperture that lie perpendicular to the incident polarization indicate that the aperture can support plasmonic excitations. The comparable near-field and focused polar Kerr signals suggest that such plasmonic excitations may lead to an enhanced near-field MOKE. This work demonstrates that near-field TRSKM can be performed without significant diminution of the polar Kerr signal in relatively large, sub-wavelength diameter apertures, while development of a near-field AFM probe utilizing plasmonic antennas specifically designed for measurements deeper into the nanoscale are discussed.

# I. INTRODUCTION

Time-resolved scanning Kerr microscopy (TRSKM) has been used extensively for the investigation of picosecond magnetization dynamics at sub-micron lengthscales. Since 1996, TRSKM has been adopted as a characterization tool for a broad range of dynamic magnetic phenomena in metallic thin film ferromagnets for spintronic, magnonic, and data storage applications. TRSKM was first used for the imaging of the flux dynamics of hard disk thin film write heads[1,2] and magnetization precession in a microscale disc[3]. Shortly after, vector-resolved precessional motion[4], ultrafast reversal[5], and precessional switching[6,7] were reported in anticipation of such dynamics being incorporated into hard disk sensor technology and magnetic random access memory elements (MRAM). In a microscale stripe TRSKM was used to image so-called center and edge localized spin wave modes[8], while in a microscale square a change in spin wave relaxation was observed[9], where both phenomena were related to the particular configuration of the equilibrium magnetization. At the same time, TRSKM utilizing microwave excitation was developed allowing individual, resonantly excited, spin wave modes to be imaged[10,11].

These early studies directly imaged the spatial character of the dynamic magnetization in microscale thin film structures. However, TRSKM could also be used to detect localized spin waves in nanoscale elements, both within arrays[12-14], and within individual nanomagnets[15-17] that were smaller than the focused spot of the optical probe and more applicable to sensor and MRAM technologies. In relation to the emerging field of magnonics, TRSKM was used to image collective spin wave excitations in microscale arrays of dynamically dipolar coupled nanomagnets[18], and later to isolate the effect of the dynamic interaction on the spin wave resonances of a pair of nanoscale discs by probing each disc individually[19]. TRSKM has also been used to image propagating spin waves generated by a variety of excitations[20-23] for potential magnonic applications.

The flexibility of TRSKM has also permitted dynamical studies of magnetic vortices[24-29], rings[30,31], domain walls[32], microwave assisted magnetic switching[33], and hard disk writers for perpendicular media disks[34-36]. In the last decade, emerging phenomena in spintronics have also been explored using TRSKM. These include spin-pumping in magnetic heterostructures[37,38], and the influence of non-local spin injection[39], the spin Seebeck effect[40], and the Dzyaloshinskii–Moriya interaction[41]. Some of the most recent applications of TRSKM to spintronics have revealed the spatio-temporal character of magnetization dynamics generated by autonomous vortex gyration in nano-contact spin torque oscillators[42-44], and spin-orbit torque switching[45,46].

In these studies using TRSKM, the magnetization dynamics are excited by either pulsed or microwave magnetic fields, an optical pulse, or by spin transfer or spin-orbit torques. The dynamics are then detected by stroboscopic measurement of the magneto-optical Kerr effect (MOKE) using ultrafast laser pulses that are synchronous with the excitation. Since TRSKM is a scanning laser microscopy technique, its spatial resolution is determined by the diffraction limit of light, and only the spatially averaged dynamic magnetization within the laser spot can be detected. For the best free space spatial resolution, microscope objective lenses with high numerical apertures of typically ~0.65 (×40) to ~0.8 (×60) are used. The diffraction limited spatial resolution can be further enhanced by using shorter wavelengths, typically generated by second harmonic generation using laser pulses of a modelocked Ti:sapphire laser that are often used for TRSKM[4,16]. The resulting spatial resolution is typically 600 nm and 300 nm for probe pulses with wavelengths of 800 nm and 400 nm respectively. The spatial character of the magnetization dynamics is convolved with the Gaussian profile of the focused laser spot. The lengthscale of the magnetization dynamics such as localized or propagating spin waves, and of magnetic textures such as domain walls, vortices, and Skyrmions, can be an order of magnitude smaller than the spatial resolution of TRSKM. Therefore a new approach is required to probe the spatial character of magnetization dynamics and textures beyond the diffraction limit of light, while maintaining the picosecond temporal resolution of TRSKM. Here we present a TRSKM platform that overcomes this limitation by incorporating scanning near-field optical microscopy into the measurement set-up.

Scanning near-field optical microscopy (SNOM) has become a well-established technique for sub-wavelength imaging since its first demonstration in 1984[47,48]. However, only a few studies utilizing scanning near-field magneto-optical imaging have been reported in that time. Eight years after SNOM was first demonstrated, Betzig *et al.*[49] used a tapered optical fiber[50] to acquire near-field magneto-optical Faraday effect images in transmission. Images of the perpendicularly magnetized domains of a Co/Pt multilayer film demonstrated a spatial resolution of ~30 to 50 nm. Two years later, Silva *et al.*[51] used SNOM to image magnetic domains in optically opaque materials. Their approach used the optically excited surface plasmon resonance of a 30-nm Ag nano-particle as a near-field probe. By analyzing the polarization rotation of scattered light due to the MOKE, they were able to image in reflection. Again, images of the domains of a Co/Pt multilayer film were acquired with a spatial resolution of better than 100 nm. In general, magneto-optical SNOM in reflection is advantageous since magnetic devices and integrated waveguides for TRSKM are often opaque, while at the same time any substrate can be used.

In the year following the first report of TRSKM by Freeman *et al.*[1], the same group demonstrated the enhanced spatial resolution of the technique beyond the diffraction limit[52]. Rather than replacing the high numerical aperture microscope objective lens with a tapered optical fiber or plasmonic nanoparticle, they instead placed a solid immersion lens (SIL) beneath the objective lens and in close proximity to the sample (within an optical wavelength). The evanescent optical near-field generated at the base of the SIL at the point of total internal reflection provided sensitivity to the out-of-plane component of the magnetization using polar MOKE. Time-resolved spatial profiles of magnetization reversal in the pole tip of a magnetic recording head were acquired with a spatial resolution of ~200 nm using a truncated-SIL. More recently, in 2015, Rudge *et al.*[53] reported the use of a tapered optical fiber to detect the evanescent optical field within 100 nm of the surface of a 1 µm-diameter CoFeB disc. The disk was uniformly illuminated at an acute angle of incidence using a low numerical aperture microscope objective lens allowing sensitivity to the in-plane component of the magnetization using the longitudinal MOKE. The near-field approach was also adopted by Jersch *et al.*[54] in 2010 for Brillouin light scattering (BLS) experiments that, in contrast to TRSKM, are conducted in the frequency domain. In their approach an atomic force microscope was integrated with the BLS apparatus and used to position a hollow pyramidal Al tip with a 50 nm aperture in its apex approximately 50 nm from the sample surface. The high sensitivity of the BLS technique allowed the precise spatial character of edge modes in a microscale NiFe ellipse to be characterized with a spatial resolution of ~50 nm. While near-field techniques have been successfully applied to studies of magnetization dynamics using both TRSKM and BLS, their lack of continued use and the lack of a well-established measurement platform imply that the use of such techniques for routine measurements is challenging. For example, the optical transmission of apertures with sub-wavelength diameter is proportional to $(r/\lambda)^4$ where $r$ and $\lambda$ are the radius of the aperture and the wavelength of the light respectively.[55] Such diminution of the transmission can necessitate infeasibly large image acquisition times during which thermally activated mechanical drift over micrometer lengthscales can compromise image quality.

In this work we demonstrate a platform for near-field TRSKM using a nanoscale metallic aperture fabricated on the tip of an optically transparent atomic force microscope (AFM) probe, Figure 1(a). The aperture was positioned in close proximity to the sample surface (within one diameter) and images were acquired in reflection. The electric near-field of the aperture could then be considered as a coupled mode with the sample surface to overcome the rapid diminution of transmission for sub-wavelength apertures. The AFM was mounted onto the microscope column, replacing the conventional high numerical aperture microscope objective lens typically used for TRSKM, Figure 1(b). Time-resolved Kerr images of the dynamic response of a 5×2.5 µm² NiFe(40 nm) rectangle with a flux-closure equilibrium state were acquired at remanence. The images acquired using the near-field AFM tip have been compared to those acquired with a ×60 microscope objective lens. The spatial resolution of the near-field Kerr images was enhanced with respect to images acquired with the objective lens, revealing evidence of magnetization dynamics with finer spatial character that may be associated with the domain walls of the flux-closure state. The diminution of the near-field spatial resolution as a function of tip-sample separation was found to be consistent with a numerical model of the spatial profile of the electric field generated between a metallic aperture and the surface of a ferromagnetic metal.

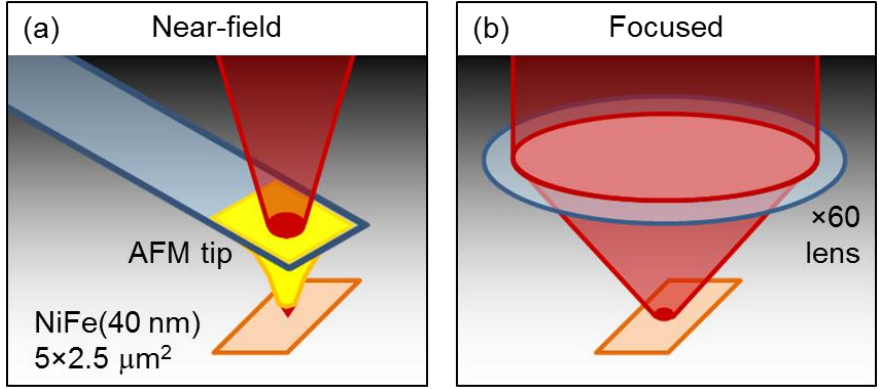

Figure 1. Illustrations of the optical probe used in (a) near-field Kerr microscopy using a nanoscale aperture fabricated on the tip of an AFM probe, and (b) focussed Kerr microscopy using a microscope objective lens.

## II. EXPERIMENTAL SET-UP AND SAMPLE DETAILS

*A. Time-resolved scanning Kerr microscopy*

Time-resolved scanning Kerr microscopy was performed using the laser pulses from an ultrafast Ti:Sapphire oscillator that were generated with sub-100 fs duration, 80 MHz repetition rate, and 800 nm wavelength, Figure 2. The laser repetition rate was synchronized to an 80 MHz trigger waveform of an external oscillator (clock) that was also used to trigger an impulse generator at the same rate). The output of the impulse generator was gated at ~3 kHz for amplitude modulation of the impulse waveform, which was then passed through a coplanar waveguide (CPW). The in-plane component of a magnetic field pulse $h_x(t)$ associated with the current waveform in the CPW was used to excite magnetization dynamics of the NiFe rectangle that was fabricated directly on top of the CPW center conductor, Figure 2(a). Time-resolved Kerr images were acquired in the absence of a bias magnetic field so that a flux-closure domain structure of the in-plane magnetization was formed within the rectangle, as confirmed by magnetic force microscopy[56] (MFM) of Figure 2(b), and illustrated by the adjacent sketch.

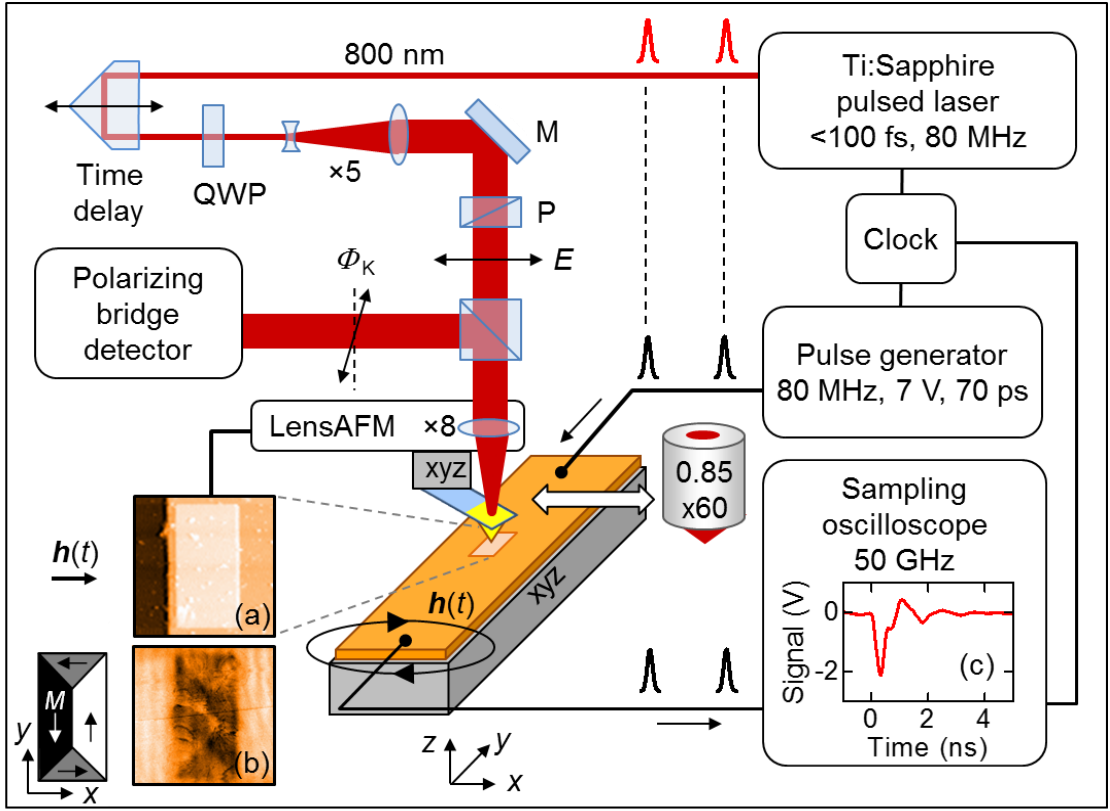

Figure 2. A schematic of the experimental apparatus for near-field TRSKM. The LensAFM can be interchanged (indicated by white block arrow) with a conventional high numerical aperture lens (0.85, ×60) for focused TRSKM measurements. Inset images of topography (a) and phase contract (b) acquired from the 5×2.5 μm² rectangle using the LensAFM for contact mode AFM and magnetic force microscopy (MFM) respectively. A sketch of the flux-closure domain structure of the in-plane equilibrium magnetization is shown adjacent to the MFM image.

The 5×2.5 μm² NiFe(40 nm) rectangle (the rectangle) was fabricated on top of the center conductor of a coplanar waveguide (CPW) that was used to excite magnetization dynamics within the rectangle. The rectangle and CPW were fabricated from a Ta(5)/[Cu(25)/Ta(3)]$_{\times 3}$/Cu(25)/Ta(10)/Ru(5)/Ni$_{81}$Fe$_{19}$(40)/ Al(1.5) multilayer that was deposited by DC magnetron sputtering onto a Sapphire substrate with thickness of 500 μm. Self-passivation of the Al layer by air oxidation provided a protective cap for the ferromagnetic NiFe layer. The NiFe layer was first patterned into magnetic elements using a combination of electron beam lithography and ion beam milling down to the Ru layer. Photolithography was then used to pattern the exposed Cu/Ta underlayers into a CPW structure such that the NiFe elements were arranged along its center conductor. The width of the CPW center conductor and its separation from the ground planes were 6 μm and 3.75 μm respectively. The CPW was designed for a characteristic impedance of 50Ω, which allowed good transmission of the impulse waveform, Figure 2(c). The thickness of the CPW structure (~ 130 nm) and the NiFe(40 nm) elements was confirmed by cross-sections (not shown) extracted from AFM topography images of the sample, Figure 2(a). The phase-contrast MFM image (Figure 2(b)) reveals contrast that is expected of a flux-closure domain configuration of the in-plane equilibrium magnetization that supports a cross-tie domain wall along the center of the rectangle.[57] It should be noted that MFM is sensitive to the stray magnetic field of the rectangle, and not the equilibrium magnetization. For clarity, a sketch of the flux-closure domain structure of the in-plane equilibrium magnetization is shown adjacent to the MFM image.

For diffraction limited (focused) TRSKM measurements the pulsed laser beam was expanded ×5 to reduce the beam divergence and then linearly polarized (×5 and P, Figure 2). The beam was then focused directly onto the sample surface using a ×60 (0.85 numerical aperture) microscope objective lens. The rear aperture of the objective lens was completely illuminated by the incident beam. In near-field TRSKM measurements the objective lens was replaced with a Nanosurf LensAFM. The AFM was equipped with a ×8 objective lens that was used to focus the pulsed laser beam onto the detector side of the AFM cantilever directly above the tip. Near-field Kerr images were acquired for two different probe polarizations, parallel and perpendicular to the long edge of the rectangle. To maintain the incident intensity of the beam for the different polarizations, a quarter-wave plate was inserted into the beam upstream of the beam expander (QWP, Figure 2). Fine adjustment of the intensity was achieved using a neutral density filter wheel located immediately after the exit aperture of the beam expander (not shown).

At a fixed position on the rectangle time-resolved Kerr traces were acquired by scanning a retroreflector to change the time-delay between the pulsed magnetic field and the probe laser pulse. Then, at a fixed time-delay time-resolved Kerr images were acquired by scanning the sample beneath the focused or near-field optical probes using a piezoelectric flexure stage. Images were acquired by scanning along the *y*-direction (fast axis) for each position in the *x*-direction (slow axis). In both the near-field and focused Kerr images the step size along the *x*- and *y*-direction was 200 nm. Amplitude modulation of the pulsed magnetic field generated by the CPW was used to modulate the excitation of the magnetization. In turn the modulated out-of-plane component of the magnetization modulated the reflected polarization due to the polar MOKE. The amplitude and phase of the polar Kerr signal were then recovered at the modulation frequency using a polarizing optical bridge detector and a lock-in amplifier.

*B. Near-field tip fabrication*

In this work, two similar AFM probes were used for near-field TRSKM and *in-situ* contact-mode AFM. The cantilever and the tip of both probes were made of an undisclosed optically transparent quartz-like material. To minimize damage to the nanoscale aperture, commercially available NanoWorld qp-CONT AFM probes (with sharp tips) were used to locate the rectangle using contact-mode AFM. For near-field TRSKM measurements, prototype NanoWorld qp-CONT type probes with plateau tips were used. The plateau provided a flat surface for the fabrication of the nanoscale aperture on the underside of the tip, while the optically transparent material allowed the pulsed laser beam to pass through the cantilever and tip to the aperture for near-field TRSKM. While the near-field AFM probe could also be used in contact-mode, the spatial resolution of its plateau tip was inferior to that of the sharper commercially available tips. The detector side of both probe cantilevers required minor modification for their operation since neither were specifically designed for use in the LensAFM. For the commercially available qp-CONT probes, the cantilevers were ~125 μm long, ~35 μm wide, and 750 nm thick. The tip had a trumpet-shaped profile, Figure 3(a). The force constant and resonance frequency was 0.1 N/m and 30 kHz respectively. The cantilever thickness of the prototype probe was ~1400 nm.

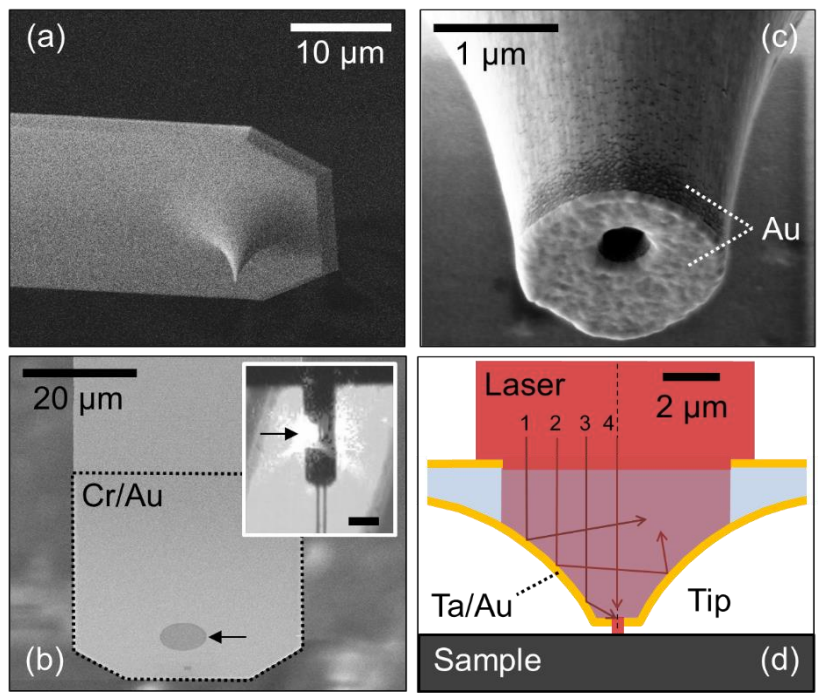

Figure 3. (a) A scanning electron microscope image of commercially available NanoWorld qp-CONT AFM probes (with sharp tip). (b) The detector side of a prototype NanoWorld qp-CONT type probe with plateau tip. An aperture with diameter of 8 μm was FIB milled into the Cr/Au pad directly above the tip (indicated by arrow). Inset in (b) shows the scatter of the LensAFM laser approximately half way along the cantilever (indicated by arrow, scale bar 20 μm). (c) The plateau tip coated with Au(200 nm) with a 400 nm diameter FIB-milled aperture close to the center of the plateau. (d) A schematic cross-section of the tip showing ray tracings of beam paths that pass through the rear microscale aperture far from the optical axis (1), closer to the optical axis (2, 3), and along the optical axis (4).

The alignment of the LensAFM laser onto the detector side of a cantilever was fixed. The position of the laser spot on the cantilever relied on mating alignment grooves of the AFM probe with those of an alignment chip in the LensAFM scan head. For qp-CONT probes not designed for use with the LensAFM, the laser was not positioned on the reflective Cr/Au(60 nm) square pad at the free end of the cantilever, directly above the tip where the deflection was largest, Figure 3(b). Instead the laser was positioned approximately half way along the optically transparent cantilever, Figure 3(b) (inset, indicated by arrow). To prevent transmission of the LensAFM laser through the cantilever and loss of cantilever deflection signal, Pt containing material was deposited[58] over the uncoated region of the detector side of the cantilever using an FEI DualBeam focused ion beam - scanning electron microscope (FIB-SEM) system. The Pt containing material significantly enhanced the reflectivity for the LensAFM laser so that the cantilever deflection signal was recovered. A laser power meter revealed that the transmission of the LensAFM laser through qp-CONT probe cantilevers was reduced from 180 µW to 2 µW when Pt containing material was deposited onto the detector side of the cantilever. Since the LensAFM laser was aligned only part way along the qp-CONT cantilevers, the resulting deflection signal was small with respect to that when the laser was aligned closer to the tip of much longer cantilevers. Therefore, to enhance the sensitivity to the smaller deflection of the qp-CONT cantilevers, a calibration was applied to reduce the expected deflection by a factor corresponding to the ratio of the qp-CONT cantilever length (~125 µm) to that of AFM probes typically recommended for use in the LensAFM (~450 µm). Such modification to the qp-CONT cantilevers and the scan head set-up permitted the use of the qp-CONT probe for contact-mode AFM in the LensAFM.

Additional processing of the prototype qp-CONT plateau tip for near-field TRSKM was also required. First, Pt containing material was FIB deposited onto the detector side of the cantilever (as described above). A Ta(<10 nm seed)/Au(~180 nm) film was then deposited using DC magnetron sputtering (from a base pressure of ~1×10$^{-7}$ Torr) onto the tip and the underside of the cantilever for fabrication of the nanoscale aperture. Next, an aperture with diameter of 400 nm was FIB milled through the Ta/Au film at the centre of the tip plateau to expose the quartz-like tip material, Figure 3(c). The acceleration voltage and beam current were 30 kV and 50 pA respectively. Finally, a microscale hole with diameter of 8.5 µm was FIB milled through the Cr/Au square cap on the detector side of the cantilever to expose the cantilever, Figure 3(b). Since the cantilever and tip were optically transparent, it was not necessary to FIB mill through the entire tip. The pulsed laser beam for TRSKM was focused using the ×8 objective lens of the LensAFM to form a spot that overfilled the microscale hole in the Cr/Au cap directly above the tip. The beam then passed through the cantilever and tip to the nanoscale aperture. The trumpet-shaped tip plays a role in guiding the pulsed laser beam down to the nanoscale aperture, Figure 3(d). Rays far from the optical axis may be reflected back out of the tip by single (1) or multiple reflections (2). In contrast, rays close to the optical axis (3) or propagating along it (4) may be reflected towards the nanoscale aperture or pass directly to it, leading to the excitation of localized surface plasmons on the perimeter of the aperture[59] To prevent scattered pulsed laser light from entering the LensAFM photodetector, a bespoke short-pass 731 IK dichroic filter was mounted in front of the LensAFM photodetector by the manufacturer Nanosurf. The transmission at the wavelengths of the LensAFM detection laser (650 nm) and the TRSKM pulsed laser (800 nm) was >85% and ~1%, respectively.

*C. Near-field image acquisition*

To locate the rectangle for focused TRSKM it was straightforward to scan the sample over a large area and identify the element from an image of the reflectivity. For near-field TRSKM the rectangle was first located using contact-mode AFM with the modified qp-CONT probe. Once the rectangle was located, the AFM image offset was noted and the tip retracted to its home position, before removing the LensAFM from a quick release mount attached to the TRSKM microscope column. The contact-mode AFM tip was then removed and replaced with the near-field TRSKM tip, and the LensAFM returned to the microscope column. On approach to the sample surface the image offset noted from the AFM images was applied. The tip repositioning accuracy of the LensAFM alignment system was typically better than ± 8 µm. Therefore, both contact-mode AFM and near-field TRSKM were performed using the same type of AFM probes so that the position of the rectangle located using contact-mode AFM would be close to the near-field tip. This protocol avoided the need to scan large areas with the near-field tip, which could potentially degrade the nanoscale aperture due to the large topography of the CPW. The location of the rectangle was then identified by first setting the retroreflector to a position corresponding to a time-delay identified from focused TRSKM to be after the arrival of the pulsed magnetic field, i.e. a time delay at which a dynamic polar Kerr signal could be expected. Then the sample was scanned beneath the near-field tip to search for a polar Kerr signal. Once the Kerr signal was detected, it was then optimized by using mirror M (Figure 2) to adjust the position of the incident pulsed laser beam on the microscale hole in the Cr/Au cap directly above the tip.

Near-field Kerr images were acquired by first approaching the surface of the rectangle until contact. When the tip was in contact with the sample, the near field aperture remained a small distance from the surface of the rectangle due to the approach angle of ~9° between the sample surface and AFM probe cantilever. The approach angle ensured sufficient clearance between the surface and the AFM probe chip attached to the cantilever so that only the tip made contact with the surface. Furthermore, the approach angle only permitted the leading edge of the plateau to make contact with the surface. When the leading edge of the plateau was in contact with the surface, the centre of the aperture was calculated to be ~135 nm from the surface for an approach angle of 9° and plateau radius of 0.86 µm, corresponding to ~$d$/3 where $d$ is the aperture diameter. . Once in contact, the near-field tip was then made idle, but in a closed-loop configuration (unless otherwise stated) in which the deflection of the cantilever was monitored. The sample was then scanned beneath the near-field tip at a rate of at least ~1.5 s/pixel to accommodate a 500 ms lock-in integration time for the dynamic polar Kerr signal. To monitor tip-sample contact during the image acquisition, the cantilever deflection signal was acquired simultaneously along with reflectivity, time-averaged polar Kerr signal, and dynamic polar Kerr signals of the near-field TRSKM. The contact between the leading edge of the tip and the sample ensured a constant tip-sample separation of ~$d/3$ during the image acquisition.

## III. RESULTS AND DISCUSSION

*A. Near-field time-resolved scanning Kerr images*

*i. Initial response*

Focused and near-field time-resolved Kerr images acquired with optical probes linearly polarized along the *x*-direction (short edge of the rectangle) are shown in Figure 4(a) and Figure 4(b) respectively. Near-field images in Figure 4(c) were acquired with the probe polarized along the *y*-direction, and will be discussed later. The images of Figure 4(a) and Figure 4(b) have been selected from a compilation of time-resolved images that have been made into movies of the magnetization dynamics[60]. The initial motion of the antiparallel magnetization of the left and right domains was out-of-plane, but in opposite *z*-directions yielding contrasting Kerr rotations. This is expected because of the opposite torque generated by the in-plane pulsed magnetic field applied perpendicular to the equilibrium magnetization of the left and right domains (Figure 2 inset). Accordingly, the flux-closure equilibrium state provides maximum contrast of the dynamic magnetization across the domain wall positioned along the center of the rectangle in the *y*-direction. The domain wall is expected to be approximately one order of magnitude narrower than the diameter of the near-field and focused laser spot. The change in the Kerr signal across the domain wall therefore represents the convolution of the optical probe and the relatively sharp transition in the dynamic response of the magnetization in the left and right domains. Such contrast allowed a direct comparison of the magneto-optical spatial resolution using focused and near-field TRSKM.

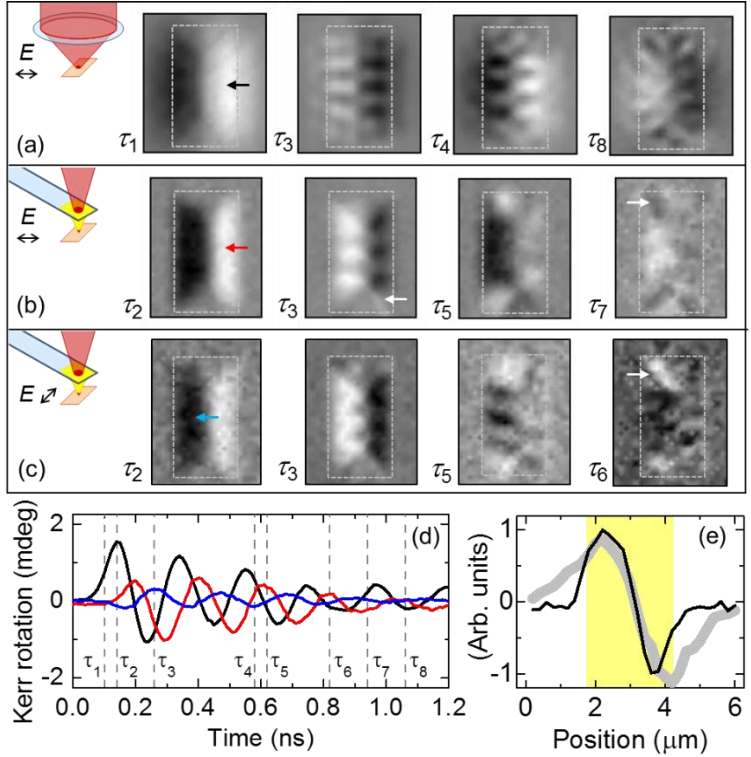

Figure 4. Focused (a) and near-field (b, c) time-resolved polar Kerr images acquired using incident polarization parallel (a, b) and perpendicular (c) to the long edge of the rectangle. The black-grey-white contrast corresponds to the change in the out-of-plane component of the dynamic magnetization, where black and white correspond to a changes of opposite sign. (d) Time-resolved traces acquired from the right (a, b) and left (c) closure domain at locations indicated by arrows of the same color in (a-c). The time-delay at which images in (a-c) were acquired are indicated by vertical dashed gray lines labelled with the corresponding time-delay of each image $\tau_i$. (e) Profiles extracted from focused and near-field Kerr images at $\tau_1$ and $\tau_2$ in (a) and (b) respectively. The profiles were extracted across the width (*x*-direction) and approximately through the center of the rectangle.

The focused and near-field Kerr images of Figure 4(a) and Figure 4(b) acquired at time-delays of $\tau_1$ and $\tau_2$ respectively (see Figure 4(d)) reveal that the initial response of the left and right domain magnetizations was quite uniform. However, it is clear that the focused Kerr image at $\tau_1$ in Figure 4(a) revealed contrast outside of the superimposed outline of the rectangle. Similarly the magnetic contrast from the left to the right domain appears to be smoothed over length scales that are larger than the expected spot size. In contrast, such artefacts were not observed in the near-field Kerr images. Instead, the image at $\tau_2$ in Figure 4(b) appears to exhibit sharper contrast, with enhanced definition of the domain walls and end closure domains (gray) along the short edges of the rectangle. Furthermore, the contrast of the near-field Kerr images was not observed to extend significantly outside of the perimeter of the rectangle. Remarkably, the near-field technique yields a substantial polar Kerr signal (~1 mdeg) that is more than half that of the focused Kerr signal (see Figure 4(d)) despite the potential loss of probe light due to internal reflections within the AFM tip (see Figure 3(d)). This suggests that the Kerr rotation originating from the near-field tip was enhanced with respect to that of the focused technique, and may be ascribed to an enhanced MOKE due to the excitation of localized surface plasmons along the perimeter of the aperture,[59] which will be discussed later (Section III.C).

Line scans extracted from the Kerr images acquired at $\tau_1$ in Figure 4(a) and and $\tau_2$ Figure 4(b), using the focused and near-field optical probes respectively, are shown in Figure 4(e). The line scans revealed a sharper transition between the positive and negative Kerr signal of the left and right domains respectively when using the near-field AFM tip (black line profile). Furthermore outside of the edges of the rectangle (indicated by yellow band in Figure 4(e)), the decrease in the Kerr signal to zero was also sharper in the near-field Kerr images, while tails of more slowly decreasing signal were observed in the line scan extracted from the focused Kerr images (grey line profile). These tails appear as contrast outside the perimeter of the rectangle in Figure 4(a) and are ascribed to spherical aberration of the microscope objective lens. The full width at half maximum (FWHM) of the focused laser spot was determined to be slightly better than ~800 nm using the knife-edge technique[61] in reflection. The spot FWHM was larger than expected and understood to be due to the spherical aberration. The aberration was identified as side lobes in the derivative of the reflectivity error function curve that the knife-edge technique generates as the spot is scanned across the sharp edge of a material with high reflectivity (not shown). The amplitude of the side lobes was less than 1/3rd that of the main Gaussian spot profile, but sufficiently large to generate the Kerr signal outside of the rectangle described above. . The improved spatial resolution and elimination of such artefacts allowed higher quality Kerr images to be acquired using the near-field AFM tip.

*ii. Relaxation*

At larger time-delays the magnetization dynamics rapidly become more non-uniform with the emergence of ripples or standing spin waves. In Figure 4(b) the contrast in the left (right) domain observed at $\tau_3$ exhibits three white (black) antinodes separated by two gray nodes. If the dynamics were standing spin waves, it would be expected that the central antinode would appear as black (white). It is known that magnetic ripples[62,63] and/or cross tie domain walls[57,64-66] can form within NiFe films with thickness similar to that of the rectangle imaged in this work and so cannot be ruled out. Indeed, the magnetic contrast of the MFM image of the rectangle in Figure 2(b) is consistent with that observed previously[57] for a similar NiFe rectangle with size (3×1 μm$^2$) and thickness (50 nm) that supports a cross-tie domain wall at its centre and along its length. The presence of a cross-tie domain wall may then encourage the formation of the observed ripple-type dynamics. The spatial character of the phase contrast observed in the MFM image of Figure1(b) is seen to be in good agreement with that of the ripple structure observed in the Kerr images. Unlike the near-field image, the focused Kerr image at $\tau_3$ in Figure 4(a) shows additional weaker antinodes at each end of the left and right domains. These are understood to be artefacts, again due to the spherical aberration ,since they are not observed in the near-field image at $\tau_3$. Instead, the near-field Kerr image reveals evidence of magnetization dynamics associated with the diagonal domain walls of the end closure domains, as indicated by the arrow in the image at $\tau_3$ in Figure 4(b). At $\tau_3$ in Figure 4(a) such detail is not clearly observed in the focused Kerr image.

At time-delays $\tau_4$ to $\tau_8$ in Figure 4(a) and Figure 4(b) the dynamics become more complicated as the relaxation continues. At $\tau_4$ the focused Kerr images (Figure 4(a)) reveal non-uniform contrast within the end closure domains and in the vicinity of the diagonal domain walls, but it is unclear if diffraction of the stronger magneto-optical signal of the left and right domains obscures the end domain dynamics. At $\tau_8$ the left and right domains exhibit more complicated dynamics with reduced amplitude, which allows the dynamics of the end domains to be observed more clearly. At $\tau_8$ in Figure 4(a), the alternating black-white-black-white contrast (from left-to-right) within the top end domain is more akin to the expected contrast of a confined spin wave. The near-field images (Figure 4(b)) show a corresponding diminution of the amplitude of the dynamics at $\tau_5$ and $\tau_7$. The near-field images also reveal asymmetry in the response of the left and right domains, and reveal dynamics in the vicinity of the diagonal walls of the end domains.

*iii. Probe polarization*

To confirm that the near-field Kerr images were indeed sensitive to the out-of-plane component of the dynamic magnetization by means of the polar Kerr effect, the polarization of the probe laser pulse was rotated through 90°. At normal incidence, the polar Kerr effect is equivalent for orthogonal polarizations, and so the near-field Kerr images of Figure 4(c) acquired at $\tau_2$, $\tau_3$, and $\tau_5$ were expected to be similar to those presented at equivalent time-delays in Figure 4(b). In general the near-field images of Figure 4(b) and Figure 4(c) are reasonably similar, despite the reduced signal level of images in Figure 4(c), as reflected in the time-resolved trace in Figure 4(d). Images at $\tau_7$ and $\tau_6$ in Figure 4(b) and Figure 4(c) were acquired approximately half a precession cycle apart (see Figure 4(d)) and confirm that the near field technique can spatially resolve the domain wall dynamics that appear most clearly as black and white diagonal stripes respectively in the top-left corner of the rectangle (indicated by arrows). Due to the different orientation of the equilibrium magnetization of the four diagonal domain walls their associated dynamics will be excited with different phase. Therefore, it is not expected that dynamics of all four walls will appear simultaneously in a time-resolved image at a particular delay.

*B. Confirming the near-field magneto-optical sensitivity*

To confirm that the near-field images were acquired using the electric near-field generated by the nanoscale aperture, polar Kerr images were acquired as a function of the tip-sample separation. Images of reflectivity, Kerr rotation, and deflection are shown in Figure 5(a, b, and c) respectively for intended tip-sample separations of 200, 400, 800, and 1200 nm. These images were acquired in an open loop configuration and were sensitive to mechanical drift away from the intended separation over the duration of the scan. For each image, the tip was first sent to the sample surface at the center of the rectangle. The tip was then moved away from the surface to the desired tip-sample separation before the sample was scanned beneath the tip for the image acquisition.

When the tip-sample separation was set to 200 nm the reflectivity and Kerr images exhibited sharp contrast at the edges of the rectangle and between ripples of the left and right domains respectively. However, the observation of the rectangle in the deflection image indicates that the tip was still in contact with the sample. Such contact may be due to either an attractive electrostatic interaction preventing the tip from moving a short distance (200 nm) away from the sample, or mechanical drift of a few hundred nanometers in the time between setting the tip-sample position, and the rectangle first moving beneath the tip (approximately 4 to 5 minutes). Despite contact between the sample and the tip, the near field aperture remained a small distance of $\sim d/3$ from the surface of the rectangle due to the approach angle of $\sim 9°$ of the cantilever as described previously (Section **II.**C). When the tip-sample separation was set to 400 nm the CPW center conductor gradually appears in the deflection image (Figure 5(c)). This is observed as the slowly increasing contrast of the left edge of the CPW center conductor as the sample was scanned along the fast axis ($y$-direction) parallel to the CPW. As the image acquisition progressed along the slow axis ($x$-direction), a further change in contrast was observed approximately two-thirds of the way across the rectangle such that the contrast of the remaining part of the image was similar to that observed when the tip and sample were in contact (*cf.* 200 nm, Figure 5(c)). This confirmed that the tip was not electrostatically coupled to the sample, but that the tip-sample separation gradually decreased during the image acquisition before the tip once again made contact with the surface. Typically, mechanical drift of a few hundred nanometers in the $z$-direction was not significant for focused TRSKM where the depth of focus for the ×60 objective lens was ~1 μm. However, for near-field TRSKM it was necessary to ensure the thermal stability of the microscope through the use of an enclosure to minimize air currents in and around the microscope. Despite setting the tip-sample separation $h$ to 200 nm and 400 nm respectively, the emergence of the rectangle in the deflection images of Figure 5(c) shows that the tip was in contact with the surface. Therefore both images have been labelled with a tip-sample separation of $d/3$ due to the approach angle of the AFM probe cantilever, where $d$ is the aperture diameter of 400 nm.

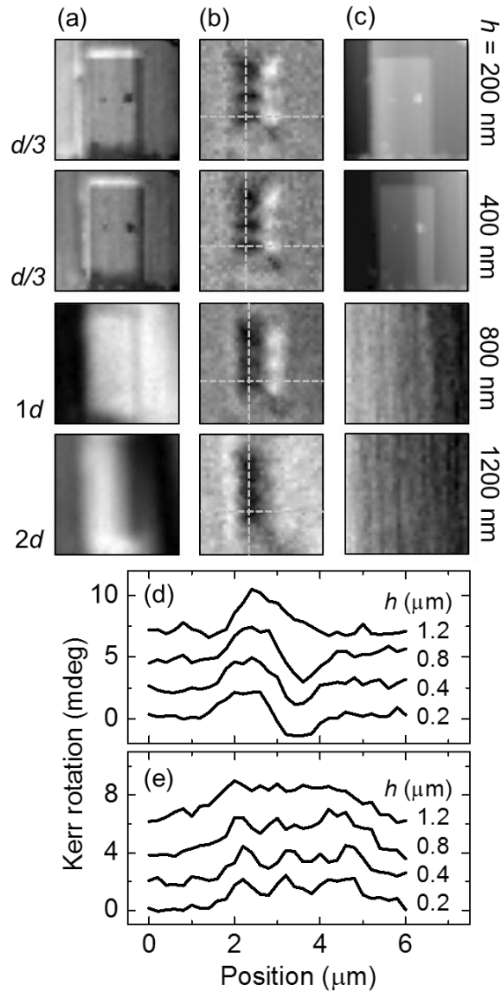

Figure 5. Reflectivity (a), polar Kerr (b), and deflection (c) images acquired using near-field TRSKM for set tip-sample separations of 200 nm, 400 nm, 800 nm, and 1200 nm. In (b) the black-grey-white contrast corresponds to the out-of-plane component of the dynamic magnetization. Lines scans extracted along the *x*- and *y*-direction from the polar Kerr images in (b) are shown in (d) and (e) respectively. The line scans were extracted along the dashed grey lines in (b).

When the tip-sample separation was set to 800 nm and 1200 nm (assumed to be $1d$ and $2d$) respectively a clear diminution of the quality of the reflectivity and Kerr images was observed (Figure 5(b and c)). At the same time no contrast was observed in the deflection images since the tip made no contact with the sample during these scans. It can be seen that the reflectivity and Kerr images became less sharp as the tip-sample separation was increased from 400 to 1200 nm. Lines scans extracted from the Kerr images in the *x*- and *y*-directions (dashed grey lines in Figure 5(b)) show the reduced sharpness more clearly, Figure 5(d and e) respectively. While it is not possible to confirm the precise tip-sample separation throughout the duration of an open-loop scan it is clear that the spatial resolution is compromised when the tip-sample separation is increased. This confirms that the reflectivity and Kerr signals are detected using the electric near-field generated by the aperture. Future modification to the near-field TRSKM scanning protocol will include a rapid approach and retract of the near-field tip at each pixel prior to the acquisition of reflectivity and Kerr signals. By identifying the location of the surface and resetting the tip-sample separation at each pixel, the negative influence of mechanical drift will be minimized. Such an operation will not significantly extend the image acquisition time, since time-resolved Kerr signals are typically integrated for 1500 to 3000 ms at each pixel.

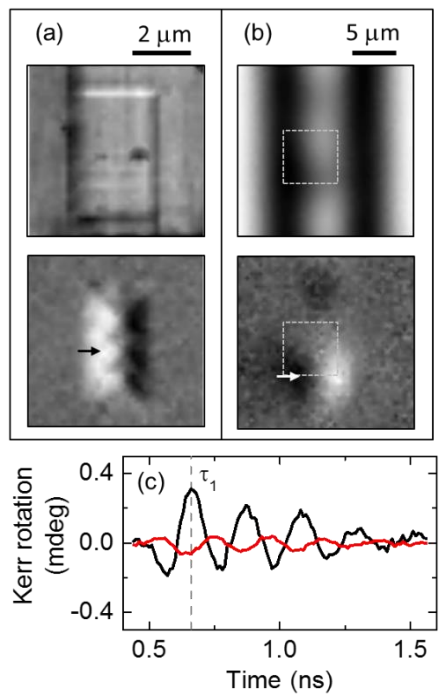

Figure 6. Images of reflectivity (top) and polar Kerr rotation (bottom) are shown for near-field TRSKM in (a) and TRSKM using the ×8 microscope objective of the LensAFM in (b). The near-field image range is indicated by the grey dashed square in (b), while the arrows in the polar Kerr images indicated the location at which the time-resolved traces in (c) were acquired.

When the near-field AFM probe was moved to the home position of the LensAFM, the pulsed laser beam was no longer focused onto the detector side of the cantilever, but instead focused onto the sample using the ×8 objective lens of the LensAFM. The depth of focus of the ×8 lens was sufficient for both the sample surface and cantilever to be considered in focus when the near-field AFM tip was a few hundred nanometers from the sample surface, Figure 2. To demonstrate that the ×8 objective lens does not provide the spatial resolution required to acquire the images shown in Figure 4 and Figure 5, reflectivity and Kerr images were also acquired using the ×8 objective lens when the near-field probe was in the home position of the LensAFM. Figure 6(a) and Figure 6(b) show reflectivity (top panel) and Kerr (bottom panel) images when using the near-field AFM tip and ×8 objective lens of the LensAFM respectively. The images were acquired at the same time delay ($\tau_1$) indicated on the time-resolved traces of Figure 6(c) that were in turn acquired at the locations indicated by arrows in the Kerr images. The probe polarization was along the *y*-direction (long edge of the rectangle).

The near-field images of Figure 6(a) were acquired first. Immediately after the AFM tip reached its home position, the images acquired using the ×8 objective of the LensAFM of Figure 6(b) were acquired with an increased scan range and laser spot diameter of ~3 μm, Figure 6(b).. The reflectivity image acquired using the ×8 objective of the LensAFM shows the ground planes and center conductor (white) of the CPW. The locations of two NiFe elements on the left edge of the center conductor can be identified from two weaker regions of white contrast, the lower of which corresponds to the 2.5×5 μm$^2$ rectangle. The position offset in the location of the NiFe rectangle is due to the slightly different optical paths when the beam passes through either the near-field aperture or the ×8 objective of the LensAFM only. Comparison of Kerr images in Figure 6(a and b) also reveals a change in sign of the Kerr rotation. The Kerr rotation is sensitive to the refractive index of the sample and the surrounding material[67]. Therefore, it is likely that, for the same change in the out-of-plane component of the magnetization, the enhanced refractive index of the quartz-like tip material and the complex refractive index of the aperture metallization yield a change in sign of the polar Kerr signal with respect to the free-space measurements using the ×8 objective of the LensAFM. Regardless of the change in sign between the two techniques, the Kerr signal remained of opposite sign in the left and right closure domains of the rectangle, and appeared to be larger than that of the focused Kerr signal. However, the large (~3 μm) spot size of the ×8 objective of the LensAFM lead to a reduced Kerr signal due to the partial convolution of the Kerr rotation of opposite sign in left and right domains. It is clear from Figure 6 that the spatial resolution observed in the reflectivity and Kerr images acquired using the near-field AFM tip is due to the electric near-field of the nanoscale aperture, and cannot be due to the focusing of the pulsed laser beam by the ×8 objective lens of the LensAFM.

*C. Finite element modelling*

To understand the operation of the near-field AFM tip finite element modelling was used to calculate the spatial character of the electric near-field generated by a nanoscale aperture. The modelling was performed using the COMSOL Multiphysics (Comsol) commercial software package. The Comsol package was chosen because the permittivity of a material can be defined as a tensor with off diagonal elements. The off-diagonal elements can then be used to describe magneto-optical effects[68] and has recently been used to understand their enhancement by nanoscale plasmonic antennas[69].

The electric field generated by a 400 nm circular aperture in a Au(200 nm) film in free space was modelled for a range of distances between the aperture film and a ferromagnetic metal. The simulated mesh had a maximum element size of 80nm. To avoid internal reflections of the electric fields non-periodic scattering boundaries and perfectly matched layers were used. The complex permittivity of the Au film at a wavelength of 800 nm was -23.971-$i$1.5010, and was obtained from permittivity data reported in [70] using interpolation algorithms of Comsol. The ferromagnetic metal was defined using the complex permittivity of -8.628 -$i$16.65 and Voigt parameter ($Q$ = -0.006 - $i$0.011) for Permalloy (Ni$_{80}$Fe$_{20}$) at a wavelength of 800 nm[71]. Magneto-optical effects were then included in the model by defining the complex permittivity of the ferromagnetic metal as

$$\varepsilon = \varepsilon_0 \varepsilon_r \begin{bmatrix} 1 & iB_xQ & -iB_yQ \\ -iB_xQ & 1 & iB_zQ \\ iB_yQ & -iB_zQ & 1 \end{bmatrix},$$

where $B_i$ are the direction cosines of the magnetization vector. The off-diagonal components of the complex permittivity tensor used to calculate the polar Kerr effect were 0.109 ±$i$0.033. Inserting these material parameters into COMSOL the polar Kerr rotation and ellipticity of a linearly polarized electric field reflected from an optically thick film of Ni$_{80}$Fe$_{20}$ at normal incidence were calculated to be ~75 mdeg, and -25 mdeg respectively, in good agreement with analytical expressions for an air/ferromagnet interface[67]. The good agreement confirms that the complex permittivity provides a reliable description of the ferromagnetic film in Comsol and will be reported in detail elsewhere[72].

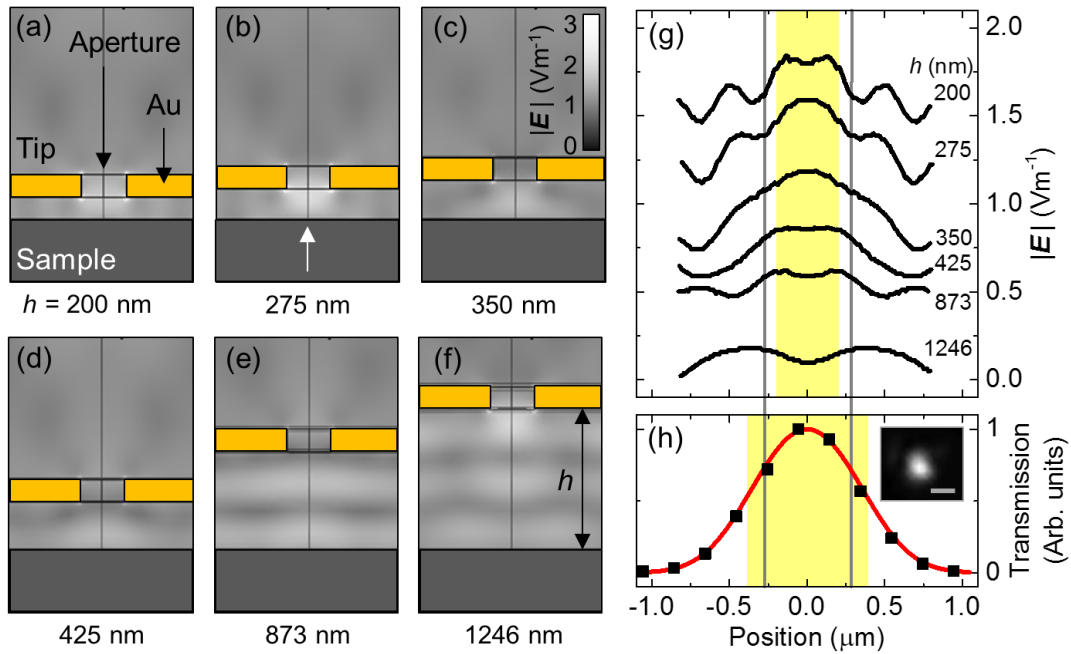

Figure 7. In (a-f) simulated spatial maps of the electric field in the vicinity of a 400-nm aperture in a Au(200 nm) film are shown for aperture-sample separations $h$ ranging from 200 nm to 1246 nm. Cross-sections in the vertical ($x$-$z$ plane) were extracted through the center of the aperture. The Au film is shown as yellow where the gap at the center of each image is the aperture. The sample is shown as dark gray. The black-gray-white contrast represents the magnitude of the electric field. In (g) profiles of the electric near-field extracted from the images in (a to f) are shown. The profiles were extracted along the $x$-direction at a distance of 50 nm from the sample surface. A cumulative offset of ~0.3 Vm$^{-1}$ has been added to each subsequent profile for clarity. (h) The transmission profile (symbols) of a 780 nm test aperture when scanned beneath the near-field AFM tip is shown overlaid with a modelled curve corresponding to a spatial resolution of 600 nm. The inset in (h) shows the near-field transmission image of the test aperture (scale bar 1 μm).

Cross-sections of the simulated electric fields for different distances (200 to 1250 nm) between the aperture and the ferromagnetic metal (sample) are shown in Figure 7(a-f). The plane of each cross-section passed through the centre of the aperture and was parallel to the linearly polarized electric field of a normally incident plane wave. The wavelength dependence of the complex permittivity can give rise to plasmonic resonances of the aperture. The linear polarization of the incident electric field has been shown in metalized tapered optical fibers to drive a spatially inhomogeneous plasmonic resonance[59]. While the 400 nm aperture used in this work was not specifically optimized for resonant excitation of such modes at a probe wavelength of 800 nm, the electric field around the perimeter of the aperture is expected to support larger amplitude electric fields in regions surrounding points on the circumference that lie perpendicular to the incident electric field. These localized near-fields can be seen as intense (~3 $Vm^{-1}$) white contrast in the immediate vicinity of the corners of the aperture cross-section on the lower side of the Au film (yellow). Towards the center of the aperture the electric field is reduced with respect to the lower corners of the aperture. While the electric field is reduced it is reasonably localized for aperture-sample distances of up to one aperture diameter (~400 nm). Line scans of the electric field extracted at a distance of 50 nm from the sample surface are shown in Figure 7(g) for the different aperture-sample distances (200 to 1250 nm). The line scans more clearly show that the electric field localization diminishes significantly as the aperture is positioned more than one aperture diameter from the sample. This is in good agreement with the diminution of the spatial resolution observed in the near-field reflectivity and Kerr images of Figure 5(a and b) and supports the interpretation that the experimental images are indeed acquired using the electric near-field of the aperture on the AFM tip.

*D. Spatial resolution*

Finally the spatial resolution of the near-field AFM tip was determined from transmission measurements on a test sample consisting of a 780 nm diameter aperture in a Au(200 nm) film. The test sample was primarily used for initial alignment of the pulsed laser beam onto the detector side of the cantilever. First, the 400-nm diameter near-field AFM tip aperture was positioned directly above the 780-nm diameter test aperture. The leading edge of the tip and the test sample were brought into contact so that the separation of the aperture from the test sample was ~$d$/3.. Next, the pulsed laser beam was aligned with the microscale rear aperture of the near-field AFM probe, directly above the tip. The alignment was optimized by monitoring the optical transmission using a photodiode directly beneath the sample. For the initial alignment only, the pulsed laser beam was chopped at ~1 kHz so that the photocurrent generated by only the transmitted beam could be recovered using a lock-in amplifier. Once the alignment was optimized a near-field transmission image of the 780 nm test aperture was acquired, see inset image of Figure 7(h).

To determine the spatial resolution of the near-field AFM tip a line scan was first extracted from the near-field transmission image across the center of the test aperture, shown as symbols in Figure 7(h). Next the convolution of a 780 nm disc with a two-dimensional Gaussian beam profile was calculated for a range of beam widths. The Gaussian beam FWHM was then varied from 100 nm to 1000 nm in 100 nm steps to identify the optimum fit to the experimental profile, Figure 7(h). For each Gaussian width, the sum of the squared difference between the points on each of the modelled convolutions and those of the experimental profile was calculated and found to be minimized for a Gaussian FWHM of ~550 nm.

The experimentally determined spatial resolution is in good agreement with the numerical modelling. For a tip-sample separations from 200 to 425 nm, the simulated profiles of Figure 7(g) extracted at a distance of 50 nm from the sample surface, show that the electric near-field is quite well confined within the experimentally determined spatial resolution of 550 nm (indicated by vertical grey lines). However, the electric field exhibits tails of reducing amplitude that extend outside of the perimeter of the AFM tip aperture. For tip-sample separations of 200 nm to 350 nm the electric field amplitude is largest on the optical axis passing through the center of the aperture. As the distance from the optical axis increases, the amplitude begins to decrease before reaching the perimeter of the aperture. At a tip-sample separation of 425 nm the electric field amplitude is almost uniform within the radius of the aperture (indicated by the yellow band), but then begins to decrease within the FWHM radius of the experimentally determined spot size.

The electric field profiles for tip-sample separations of 200 and 275 nm also reveal side lobes due to electric fields that are confined laterally from the optical axis, but also vertically between the sample and the Au film. These lobes appear in the profiles of Figure 7(g) at 400 to 500 nm from the center of the aperture for tip-sample separations of 275 and 200 nm respectively. The presence of the side lobes may compromise the near-field spatial resolution at the closest tip-sample separations of 200 and 275 nm. At a tip-sample separation of 350 nm the side lobes are no longer separated from the main peak in the electric near-field and form a broader profile that exhibits tails that extend beyond the experimentally determined FWHM radius of the experimentally determined spot size. At a tip-sample separation of 425 nm the side lobes are no longer observed.

At larger tip-sample separations of 873 and 1246 nm, the electric field exhibits local minima on the optical axis and enhanced electric field amplitude inside and outside of the tip perimeter respectively. For the larger separation the local minima becomes significant and is understood to be due to interference between the near-field of the aperture, and a confined plane wave between the Au film of the aperture and the sample. The standing wave in this region is clearly observed in Figure 7(d, e, and f) that reveal one, two, and three half-wavelengths for separations of 425, 873, and 1246 nm respectively.

## IV. SUMMARY AND OUTLOOK

In this report we have reviewed the importance of TRSKM for understanding the picosecond magnetization dynamics on sub-micron lengthscales, and recent efforts to extend its spatial resolution beyond the diffraction limit by means of near-field magneto-optical imaging. Our development of a platform for near-field TRSKM has been described in detail and we have demonstrated enhanced spatial resolution of time-resolved Kerr images acquired using near-field in comparison to those obtained by focused Kerr microscopy. We have confirmed the near-field operation by exploring the influence of the tip-sample separation, and have determined the spatial resolution to be ~550 nm for the 400 nm aperture, in good agreement with finite-element modelling. While our approach does not yet exceed the spatial resolution previously reported[52,53], it is clear that such resolution can be attained by the development of AFM probes with smaller metallic apertures. This work demonstrates that near-field TRSKM can be performed without significant diminution of the polar Kerr signal compared to far-field measurements. This experimental observation, and the prediction of enhanced electric field along regions of the perimeter of the modelled aperture that lie perpendicular to the incident electric field, suggest that localized surface plasmons may play a role in enhancing the near-field MOKE. With the realization of an enhanced near-field MOKE, measurements with smaller metallic apertures may become feasible. Further development of the near-field AFM probe is needed to extend such measurements deeper into the nanoscale. For example, development of sub-wavelength plasmonic antennas such as nanoscale crosses[73] to replace the circular aperture may yield a more localized optical near-field in close proximity to a metallic sample, but may also further enhance the MOKE in the near-field,[74] while at the same time preserve the resulting polarization changes.[73] A limiting step of our near-field platform is the individual modification of commercially available AFM probes. However, there is potential for wafer scale manufacture of near-field AFM probes for our approach, which would see near-field TRSKM become a more routinely used instrument for investigating nanoscale magnetization dynamics. Such developments will open up opportunities for time-resolved measurements of high frequency magnetic processes in magnetic textures such as vortices, domain walls and Skyrmions that are of importance for future high-frequency spintronic and magnetic storage devices.

**ACKNOWLEDGEMENTS**

The authors gratefully acknowledge the financial support of the UK Engineering and Physical Sciences Research Council under grant EP/I038470/1 'A plasmonic antenna for magneto-potical imaging at the deep nanoscale.' The authors extend special thanks to David Morgan (Windsor Scientific, UK), and Pieter van Schendel and Marco Portalupi (Nanosurf, Switzerland), for their continued advice and support for the development of the LensAFM for intergration with our TRSKM. Special thanks also go to Laure Aeschimann (NanoWorld, Switzerland) for the kind provision of prototype qp-CONT type AFM probes with plateau tips for our development of the near-field tip.